# Band gap renormalization and work function tuning in MoSe$_2$/hBN/Ru(0001) heterostructures

Qiang Zhang[1,2][†] Yuxuan Chen[1][†], Chendong Zhang[1], Chi-Ruei Pan[3], Mei-Yin Chou[3,4,5], Changgan Zeng[2,6,7][*], and Chih-Kang Shih[1][*]

[1]Department of Physics, University of Texas at Austin, Austin, TX 78712, USA
[2]Hefei National Laboratory for Physical Sciences at the Microscale (HFNL), CAS Key Laboratory of Strongly-Coupled Quantum Matter Physics, and Department of Physics, University of Science and Technology of China, Hefei, Anhui 230026, China
[3]School of Physics, Georgia Institute of Technology, Atlanta, Georgia 30332, USA
[4]Institute of Atomic and Molecular Sciences, Academia Sinica, Taipei 10617, Taiwan
[5]Department of Physics, National Taiwan University, Taipei 10617, Taiwan
[6]International Center for Quantum Design of Functional Materials (ICQD), HFNL, University of Science and Technology of China, Hefei, Anhui, 230026, China
[7]Synergetic Innovation Center of Quantum Information and Quantum Physics, University of Science and Technology of China, Hefei, Anhui, 230026, China

[†]These authors contributed equally to this work.

[*]Corresponding authors E-mail: cgzeng@ustc.edu.cn, shih@physics.utexas.edu

**Here we report the successful growth of MoSe$_2$ on single layer hexagonal boron nitride (hBN) on Ru(0001) substrate by using molecular beam epitaxy. We investigated the electronic structures of MoSe$_2$ using scanning tunneling microscopy and spectroscopy. Surprisingly, we found that the quasi-particle gap of the MoSe$_2$ on hBN/Ru is about 0.25 eV smaller than those on graphene or graphite substrates. We attribute this result to the strong interaction between hBN/Ru which causes residual metallic screening from the substrate. The surface of MoSe$_2$ exhibits Moiré pattern that replicates the Moiré pattern of hBN/Ru. In addition, the electronic structure and the work function of MoSe$_2$ are modulated electrostatically with an amplitude of ~ 0.13 eV. Most interestingly, this electrostatic modulation is spatially in phase with the Moiré pattern of hBN on Ru(0001) whose surface also exhibits a work function modulation of the same amplitude.**

The emergence of semiconducting transition metal dichalcogenides (TMDs) as two dimensional (2D) electronic materials has triggered intensive research activities[1-4]. In conjunction with graphene (a semi-metal) and hBN (a large gap insulator), they form a diverse tool set for tailoring novel 2D electronic systems. One particularly powerful approach is stacking different types of vdW materials to form vdW heterostructures[5]. Many conceptual demonstrations of vdW heterostructures have been achieved by using mechanical exfoliations of vdW layers and then stacking them together using transfer methods[6,7]. This exfoliation/transferring approach, however, is not scalable. An attractive and scalable approach is the direct epitaxial growth of 2D heterostructures, which has recently been shown in several systems using ambient chemical vapor depositions (CVDs)[8-13]. Nevertheless, achieving atomic scale control of contamination using ambient CVD is quite challenging. As an ultra-high-vacuum (UHV) based growth technique, molecule beam epitaxy (MBE) should provide better control of interface formation[14], although the number of 2D heterostructure systems demonstrated is more limited[15-19]. In this paper, we report the molecular beam epitaxial (MBE) growth and scanning tunneling microscopy/spectroscopy (STM/STS) investigation of $MoSe_2$ on single layer hBN on Ru (0001). We not only demonstrate the feasibility of growing $MoSe_2$/hBN heterostructure using MBE, but also illustrate a convincing case of electronic structure tuning of TMDs.

Single-layer hBN on transition metal surfaces, including Ru(0001), is a very interesting platform with very rich phenomena[20-22]. It had been used as a platform to grow graphene/hBN heterostructures[23-26] and moreover, such heterostructures can be separated from the Ru(0001) substrate by using electrochemical exfoliation[27,28]. Due to the slight lattice mismatch between the hBN and Ru(0001) surfaces, a so-called "nanomesh" Moire pattern forms[21,22]. Such a Moire pattern introduces not only height corrugation on hBN[21], but also periodic modulation in the local work function[29-31]. These properties make the single layer hBN/Ru(0001) and related systems an ideal platform for investigating how the local work function impacts the electronic structure of the $MoSe_2$ overlayer grown by MBE. We find the strong interaction between the hBN and Ru(0001) makes the hBN inseparable from the underlying Ru. The interaction arising from local atomic registry between the layers and the resulting charge transfer are verified by first-principles calculations. The hBN/Ru(0001) system as a whole has strong electrostatic screening effect on the band structure of $MoSe_2$, reducing the $MoSe_2$ quasi-particle gap by 0.25 eV compared to $MoSe_2$ on graphene or graphite substrates[32]. Furthermore, we show that the local work function modulation on the monolayer layer hBN results in a rigid shift of the band structure of $MoSe_2$ without changing the band gap, indicating a pure electrostatic effect. Moreover, the modulation of the $MoSe_2$ band edges is about 0.13 eV, quantitatively similar to the local work function modulation amplitude in hBN. In addition, the work function in the $MoSe_2$ shows the same modulation, illustrating a convincing case of real space electrostatic tuning of the band profile.

**Results**

**Growth and characterizations of $MoSe_2$/hBN/Ru(0001) heterostructures.**
$MoSe_2$/hBN/Ru(0001) heterostructure is synthesized in an all UHV approach. First,

single layer hBN is prepared on Ru(0001) following the standard UHV-CVD procedure[20,21]. Put briefly, hBN forms by the catalytic dehydrogenation of borazine molecules on Ru(0001) surface at proper vapor pressure and temperature. The high quality of hBN is confirmed by *in-situ* reflection high energy electron diffraction (RHEED) and STM. Shown in Fig. 1a is the RHEED pattern after the hBN growth on Ru(0001), with sharp spots indicating perfect crystallinity of the sample surface. Note that there are six dots arranged at hexagon corners surrounding each bright spot on the first Laue ring, reflecting the existence of a Moire pattern. Fig. 1b shows a typical large-scale STM image of continuous single domain hBN. A "nanomesh" moiré pattern is clearly seen. The full coverage of single-domain hBN observed here agrees with the spotty characteristic of RHEED. After a full coverage of single layer hBN, additional exposure to borazine molecule would not lead to additional growth suggesting that borazine molecules no longer have access to the catalytic Ru(0001) surface. The zoomed-in image of the nanomesh shown in Fig. 1c reveals that the periodicity of the nanomesh pattern is 3.2nm, in agreement with the periodicities of $13\times 13$ hBN and $12\times 12$ Ru(0001) lattices[21,33]. Meanwhile, two distinct topography features on the nanomesh are seen: the lower and strongly bound regions assigned as "holes," and the higher and loosely bound regions assigned as "wires." This uneven binding causes the corrugation of the hBN, with an average amplitude of about 0.1 nm.

After confirming the high quality of hBN/Ru(0001), we transferred this sample *in-situ* to the MBE system for MoSe$_2$ growth. This all-UHV approach produces a heterostructure with a clean and sharp interface. Additional sharp and uniform RHEED streaks (indicated by red arrows) in Fig. 1d reflect the successful formation of flat crystalline MoSe$_2$ layers. In Fig. 1e, the large-scale STM image shows MoSe$_2$ islands with diameters from tens to hundreds of nanometers. In Fig. 1f, the top panel shows a typical MoSe$_2$ island, the inset shows an atomic resolution image taken from single layer MoSe$_2$, and the lower panel displays the height profile along the red dashed line. From the grid in the upper panel and the height profile in the lower panel, it is evident that the superstructure of the MoSe$_2$ island has the same periodicity, corrugated amplitude and phase as the underlying hBN. Surprisingly, the expected moiré pattern of MoSe$_2$ and hBN, whose periodicity is ~1nm, is not observed here. Rather the superstructure visible on the MoSe$_2$ is just a replication of the moiré pattern from the underlying hBN, evident by the match in phase and periodicity of spatial modulation.

**Band gap renormalization of MoSe$_2$/hBN/Ru(0001) heterostructures.** In Fig. 2a we show tunneling conductance spectra with a fixed tip-sample distance, denoted as dI/dV, of MoSe$_2$/hBN/Ru(0001), together with that of MoSe$_2$/HOPG and MoSe$_2$/graphene/SiC for comparison. In these spectra, one can identify positions of the valence band state at the Γ point (labled as $E_\Gamma$), the valence band maximum (at K point), $E_V$ (located at ~ 0.4 eV above $E_\Gamma$), and the conduction band minimum, $E_C$. The quasi-particle band gap is the energy difference between $E_V$ and $E_C$. The spectra show that SL- MoSe$_2$ on graphene or on HOPG have a similar quasi-particle band gap of 2.15 eV while SL-MoSe$_2$ on hBN/Ru(0001) has a smaller quasi-particle bandgap of 1.90 eV. The results show that the quasi-particle band gap of SL-MoSe$_2$ indeed depends on the supporting substrate. Nevertheless how it is renormalized does not follow the intuition that hBN should provide a better electronic isolation of MoSe$_2$ from the substrate. Also shown in Fig. 2b is the tunneling spectrum acquired on hBN region with a relatively large sample stabilization voltage of -4 V (implying a relatively large sample-to-tip distance). Interestingly significant conductance is still present in the expected "gap region" of hBN, reflecting remanent metallic characteristics. In Fig. 2c, we show statistical distributions of the results from 102 individual tunneling spectrum measured from different locations. The average and the standard deviation of the quasi-particle band gap of the ML MoSe$_2$/hBN/Ru(0001) is 1.90 ±0.07 eV.

Thus, our study convincingly demonstrate the concept of band structure renormalization in TMDs. Nevertheless, the actual manifestation of renormalization, is probably more complex than an intuitive interpretation of the substrate electrostatic screening and warrants much more thorough investigations both experimentally and theoretically.

**Work function modulation of MoSe$_2$/hBN/Ru(0001) heterostructures.** Besides the gap renormalization discussed above, we have also observed a periodic modulation of the band profile, which is associated with the work function modulation of the nanomesh moiré pattern. Location-specific STS measurements are shown in Fig. 3. The typical dI/dV spectra taken from MoSe$_2$ hole (red curves) and wire (blue curves) regions are plotted in both the linear scale (upper panel) and the logarithmic scale (lower panel) in Fig. 3a. We use the dI/dV spectra in the logarithmic scale to determine the CBM, the VBM, and the energy of the Γ point in the valance band, $\Gamma_V$. While the band gap values and the energy differences between $\Gamma_V$ and the CBM are the same for both the hole and wire regions, there is a rigid offset for the absolute values of the CBM, VBM, and $\Gamma_V$. Such an offset in the band structure is illustrated more drastically in dI/dV mappings carried out at different tip-sample bias. Fig. 3b-c are the topography of the same area at -2.15 V and -2.0 V, which are close to $\Gamma_V$ of holes ($\Gamma_V^H$) and wires ($\Gamma_V^W$), respectively, and they look the same. However, the corresponding dI/dV mappings taken simultaneously in Fig. 3e-f have a completely reversed contrast; at the -2.15V wire regions, which are brighter in the topography images, have lower local density of state (LDOS), while at the -2V hole regions, darker in topography, have lower LDOS. This phenomena confirms the existence of the modulation of the MoSe$_2$ band structure. An alternative spectroscopy technique, tip-to-sample distance, Z, vs. bias sweep at constant current mode, is employed to determine the $\Gamma$ point values. The $(\partial Z/\partial V)_I$ spectra taken over an ensemble of 120 holes and wires (Supplementary Figure. 1.) statistically determine that, $\Gamma_V^H$ = -2.14±0.03 eV, $\Gamma_V^W$ = -2.01±0.02 eV, and therefore the amplitude of the periodic modulation of band profile is 0.13±0.05 eV.

In addition, the $(\partial Z/\partial V)_I$ spectra in the field emission regime is employed, and the sample bias for the first field emission resonance (FER) peak is considered a good approximation of the work function of the sample[30,31]. Fig. 3f-g shows the local work function modulation of hBN/Ru(0001) is 0.14 eV, and for the MoSe$_2$ overlayer such modulation is 0.16 eV, both in excellent agreement with the periodic band offset observed on MoSe$_2$. From this consistency, one can conclude that the band offset in MoSe$_2$ is purely an electrostatic effect.

**Discussion**

The bonding and resulting change in the electronic structure of hBN on Ru(0001) can be understood by first-principles calculations. The observed moiré pattern in hBN/Ru(0001) corresponds to roughly 13×13 hBN on 12×12 Ru(0001), which is too large for a thorough theoretical analysis using plane waves. Instead, we used a supercell with √57×√57 h-BN on 7×7 Ru(0001) and a small rotational angle of 6.6°, as shown in Fig. 4a, which provides the essential collection of different atomic registries between the layers and reliable electronic properties for them because the strain (0.08%) is very small. The relaxed atomic configuration is shown in Fig. 4b, indicating that a certain portion of the hBN layer is moved closer to the Ru substrate. This happens in the region near the black circle in Fig. 4a, in which the N atoms are approximately located right above the Ru atoms and strong bonding occurs, giving rise to a calculated interlayer reduction as large as 1.6 Å, in agreement with calculated results obtained previously with different supercells, basis sets, exchange-correlation functionals[34-36]. The bonding can be seen by the isosurfaces of charge transfer shown in Fig. 4c. In contrast, the green and purple circles mark the regions in which Ru atoms are located right below the B atoms and the center of the B-N hexagons, respectively. The interlayer interaction is weak, and these two regions are at about 3.7 Å above the Ru plane, a reasonable distance for the van der Waals interaction.

If we evaluate the electrostatic potential at 4.9 Å above the hBN layer (the average of the van der Waals layer separations of hBN and $MoSe_2$), the value in the hole region is clearly lower than that in the wire region by 0.1 eV, as shown in Fig. 4d. This is in excellent agreement with the relative band edge shift of SL $MoSe_2$ between the two regions as observed in the experiment. The projected density of states shown in Supplementary Figure. 2 confirms that the significant interaction between hBN and the Ru substrate in the hole region induces states in the hBN gap, giving rise to a metallic characteristics.

Our observation of a band gap reduction of 0.25 eV for MoSe$_2$ on hBN/Ru(0001) proposes interesting possibilities of band gap renormalization in 2D materials. One probable origin is the extra screening by the states in the gap of hBN arising from the strong interaction with the substrate metal. Another possible reason is the significant corrugation (about 0.1 nm) of SL MoSe$_2$ as shown in Fig. 1f. In planar SL MoSe$_2$ the VBM (CBM) is the $d_{xy}$ and $d_{x2-y2}$ ($d_{3z2-r2}$) orbitals from Mo. Any local distortion away from the perfect flatness will break the planar symmetry and induce additional hybridization between these $d$ orbitals. A band gap reduction is an entirely plausible result in this situation.

In summary, we have demonstrated the successful MBE growth of single layer MoSe$_2$ islands on top of hBN/Ru(0001). Our STM/STS results have revealed that MoSe$_2$ on the strongly coupled hBN/Ru(0001) has a quasiparticle band gap of 1.90 ± 0.07 eV, 0.25 eV smaller than the results on graphite and graphene. These results, on the one hand affirm the concept of band structure renormalization due to the substrate; but on the other hand shows that the renormalization is far more complex than a simple consideration of the metallicity of the substrate and call for more thorough theoretical/experimental investigations. In addition, we show that the local work function modulation on the hBN/Ru(0001) nanomesh structure creates a periodic template of potential modulation where the band profile of the MoSe$_2$ mimics this potential modulation precisely.

**Methods:**

**Growth of MoSe$_2$/hBN/Ru(0001) heterostructures.** The clean Ru(0001) surface was obtained on a piece of Ru single crystal by multiple cycles of Xe$^+$ sputtering and annealing at 1000 ℃. The quality of the surface was checked with STM before the Ru was transferred *in situ* to a preparation chamber, with a base pressure better than 1x10$^{-9}$ Torr. Hexagonal boron nitride formed on Ru when the single crystal was heated to 1000 ℃ and exposed to a borazine vapor of 1x10$^{-6}$ Torr for about five minutes, then the sample was slowly cooled down to room temperature. The quality of the hBN was checked by STM and RHEED. Then this hBN/Ru(0001) sample was transferred *in-situ* to a MBE chamber for the MoSe$_2$ growth. The High-purity Mo (99.95%) and Se (99.999%) elemental sources were evaporated from an e-beam evaporator and an effusion cell, respectively, with a ratio of 1:30. Single-layer MoSe$_2$ with a coverage of about 30% formed after 45 minutes of deposition with a substrate temperature of 420 ℃, followed by 30 minutes of annealing at the same temperature with the Se flux maintained.

**Scanning tunnelling microscopy and spectroscopy.** All STM investigations reported here were acquired at 77 K in UHV (base pressure is better than 6x10$^{-11}$ mbar). A tungsten tip was used. The bias voltage was applied to the sample. The conventional scanning tunneling spectroscopy, dI/dV, was acquired at a constant tip-to sample-distance (Z). The $(\partial Z/\partial V)_I$ spectrum was acquired when the tip-to-sample distance Z changes corresponding to the scanning of bias *V* in order to keep the constant current.

**First-Principles Calculations.** We have performed first-principles calculations with density functional theory (DFT) as implemented in the Vienna Ab initio Simulation Package (VASP)[37]. We used the projector augmented wave (PAW) method[38] to treat core electrons and the Perdew-Burke-Ernzerhof (PBE) form[39] for the exchange-correlation functional with a plane-wave cutoff energy of 300 eV. The periodic slabs contain three Ru layers as the substrate and a vacuum region of about 13 Å. The bottom of the three Ru layers is fixed, while the rest two Ru layers and the hBN layer are allowed to relax during the geometry optimization.

**References:**

1. Wang, Q. H.; Kalantar-Zadeh, K.; Kis, A.; Coleman, J. N.; Strano, M. S., Electronics and optoelectronics of two-dimensional transition metal dichalcogenides. *Nat. Nanotechnol.* **7**, 699-712 (2012).
2. Chhowalla, M.; Shin, H. S.; Eda, G.; Li, L.-J.; Loh, K. P.; Zhang, H., The chemistry of two-dimensional layered transition metal dichalcogenide nanosheets. *Nat. Chem.* **5**, 263-275 (2013).
3. Butler, S. Z.; Hollen, S. M.; Cao, L.; Cui, Y.; Gupta, J. A.; Gutiérrez, H. R.; Heinz, T. F.; Hong, S. S.; Huang, J.; Ismach, A. F.; Johnston-Halperin, E.; Kuno, M.; Plashnitsa, V. V.; Robinson, R. D.; Ruoff, R. S.; Salahuddin, S.; Shan, J.; Shi, L.; Spencer, M. G.; Terrones, M.; Windl, W.; Goldberger, J. E., Progress, Challenges, and Opportunities in Two-Dimensional Materials Beyond Graphene. *ACS Nano* **7**, 2898-2926 (2013).
4. Splendiani, A.; Sun, L.; Zhang, Y.; Li, T.; Kim, J.; Chim, C.-Y.; Galli, G.; Wang, F., Emerging Photoluminescence in Monolayer MoS2. *Nano Lett.* **10**, 1271-1275 (2010).
5. Geim, A. K.; Grigorieva, I. V., Van der Waals heterostructures. *Nature* **499**, 419-425 (2013).
6. Haigh, S. J.; Gholinia, A.; Jalil, R.; Romani, S.; Britnell, L.; Elias, D. C.; Novoselov, K. S.; Ponomarenko, L. A.; Geim, A. K.; Gorbachev, R., Cross-sectional imaging of individual layers and buried interfaces of graphene-based heterostructures and superlattices. *Nat. Mater.* **11**, 764-767 (2012).
7. Kretinin, A. V.; Cao, Y.; Tu, J. S.; Yu, G. L.; Jalil, R.; Novoselov, K. S.; Haigh, S. J.; Gholinia, A.; Mishchenko, A.; Lozada, M.; Georgiou, T.; Woods, C. R.; Withers, F.; Blake, P.; Eda, G.; Wirsig, A.; Hucho, C.; Watanabe, K.; Taniguchi, T.; Geim, A. K.;

Gorbachev, R. V., Electronic Properties of Graphene Encapsulated with Different Two-Dimensional Atomic Crystals. *Nano Lett.* **14**, 3270-3276 (2014).
8. Behura, S.; Nguyen, P.; Che, S.; Debbarma, R.; Berry, V., Large-Area, Transfer-Free, Oxide-Assisted Synthesis of Hexagonal Boron Nitride Films and Their Heterostructures with MoS2 and WS2. *J. Am. Chem. Soc.* **137**, 13060-13065 (2015).
9. Shi, J.; Liu, M.; Wen, J.; Ren, X.; Zhou, X.; Ji, Q.; Ma, D.; Zhang, Y.; Jin, C.; Chen, H.; Deng, S.; Xu, N.; Liu, Z.; Zhang, Y., All Chemical Vapor Deposition Synthesis and Intrinsic Bandgap Observation of MoS2/Graphene Heterostructures. *Adv. Mater.* **27**, 7086-7092 (2015).
10. Wang, S.; Wang, X.; Warner, J. H., All Chemical Vapor Deposition Growth of MoS2:h-BN Vertical van der Waals Heterostructures. *ACS Nano* **9**, 5246-5254 (2015).
11. Yan, A.; Velasco, J.; Kahn, S.; Watanabe, K.; Taniguchi, T.; Wang, F.; Crommie, M. F.; Zettl, A., Direct Growth of Single- and Few-Layer MoS2 on h-BN with Preferred Relative Rotation Angles. *Nano Lett.* **15**, 6324-6331 (2015).
12. Fu, L.; Sun, Y.; Wu, N.; Mendes, R. G.; Chen, L.; Xu, Z.; Zhang, T.; Rümmeli, M. H.; Rellinghaus, B.; Pohl, D.; Zhuang, L., Direct Growth of MoS2/h-BN Heterostructures via a Sulfide-Resistant Alloy. *ACS Nano* **10**, 2063-2070 (2016).
13. Chiu, M.-H.; Zhang, C.; Shiu, H.-W.; Chuu, C.-P.; Chen, C.-H.; Chang, C.-Y. S.; Chen, C.-H.; Chou, M.-Y.; Shih, C.-K.; Li, L.-J., Determination of band alignment in the single-layer MoS2/WSe2 heterojunction. *Nat. Commun.* **6**, 7666 (2015).
14. Koma, A.; Yoshimura, K., Ultrasharp interfaces grown with Van der Waals epitaxy. *Surf. Sci.* **174**, 556-560 (1986).
15. Ugeda, M. M.; Bradley, A. J.; Shi, S.-F.; da Jornada, F. H.; Zhang, Y.; Qiu, D. Y.; Ruan, W.; Mo, S.-K.; Hussain, Z.; Shen, Z.-X.; Wang, F.; Louie, S. G.; Crommie, M. F., Giant bandgap renormalization and excitonic effects in a monolayer transition metal dichalcogenide semiconductor. *Nat. Mater.* **13**, 1091-1095 (2014).
16. Bradley, A. J.; M. Ugeda, M.; da Jornada, F. H.; Qiu, D. Y.; Ruan, W.; Zhang, Y.; Wickenburg, S.; Riss, A.; Lu, J.; Mo, S.-K.; Hussain, Z.; Shen, Z.-X.; Louie, S. G.; Crommie, M. F., Probing the Role of Interlayer Coupling and Coulomb Interactions on Electronic Structure in Few-Layer MoSe2 Nanostructures. *Nano Lett.* **15**, 2594-2599 (2015).
17. Miwa, J. A.; Dendzik, M.; Grønborg, S. S.; Bianchi, M.; Lauritsen, J. V.; Hofmann, P.; Ulstrup, S., Van der Waals Epitaxy of Two-Dimensional MoS2–Graphene Heterostructures in Ultrahigh Vacuum. *ACS Nano* **9**, 6502-6510 (2015).
18. Liu, X.; Balla, I.; Bergeron, H.; Campbell, G. P.; Bedzyk, M. J.; Hersam, M. C., Rotationally Commensurate Growth of MoS2 on Epitaxial Graphene. *ACS Nano* **10**, 1067-1075 (2016).
19. Zhang, C.; Chen, Y.; Huang, J.-K.; Wu, X.; Li, L.-J.; Yao, W.; Tersoff, J.; Shih, C.-K., Visualizing band offsets and edge states in bilayer-monolayer transition metal dichalcogenides lateral heterojunction. *Nat. Commun.* **7**, 10349 (2016).
20. Corso, M.; Auwärter, W.; Muntwiler, M.; Tamai, A.; Greber, T.; Osterwalder, J., Boron Nitride Nanomesh. *Science* **303**, 217-220 (2004).
21. Goriachko, A.; He; Knapp, M.; Over, H.; Corso, M.; Brugger, T.; Berner, S.; Osterwalder, J.; Greber, T., Self-Assembly of a Hexagonal Boron Nitride Nanomesh on

Ru(0001). *Langmuir* **23**, 2928-2931 (2007).
22. Laskowski, R.; Blaha, P.; Gallauner, T.; Schwarz, K., Single-Layer Model of the Hexagonal Boron Nitride Nanomesh on the Rh(111) Surface. *Phys. Rev. Lett.* **98**, 106802 (2007).
23. Sutter, P.; Cortes, R.; Lahiri, J.; Sutter, E., Interface Formation in Monolayer Graphene-Boron Nitride Heterostructures. *Nano Lett.* **12**, 4869-4874 (2012).
24. Roth, S.; Matsui, F.; Greber, T.; Osterwalder, J., Chemical Vapor Deposition and Characterization of Aligned and Incommensurate Graphene/Hexagonal Boron Nitride Heterostack on Cu(111). *Nano Lett.* **13**, 2668-2675 (2013).
25. Liu, M.; Li, Y.; Chen, P.; Sun, J.; Ma, D.; Li, Q.; Gao, T.; Gao, Y.; Cheng, Z.; Qiu, X.; Fang, Y.; Zhang, Y.; Liu, Z., Quasi-Freestanding Monolayer Heterostructure of Graphene and Hexagonal Boron Nitride on Ir(111) with a Zigzag Boundary. *Nano Lett.* **14**, 6342-6347 (2014).
26. Yang, W.; Chen, G.; Shi, Z.; Liu, C.-C.; Zhang, L.; Xie, G.; Cheng, M.; Wang, D.; Yang, R.; Shi, D.; Watanabe, K.; Taniguchi, T.; Yao, Y.; Zhang, Y.; Zhang, G., Epitaxial growth of single-domain graphene on hexagonal boron nitride. *Nat. Mater.* **12**, 792-797 (2013).
27. Lu, J.; Zhang, K.; Feng Liu, X.; Zhang, H.; Chien Sum, T.; Castro Neto, A. H.; Loh, K. P., Order–disorder transition in a two-dimensional boron–carbon–nitride alloy. *Nat. Commun.* **4**, 2681 (2013).
28. Sutter, P.; Huang, Y.; Sutter, E., Nanoscale Integration of Two-Dimensional Materials by Lateral Heteroepitaxy. *Nano Lett.* **14**, 4846-4851 (2014).
29. Dil, H.; Lobo-Checa, J.; Laskowski, R.; Blaha, P.; Berner, S.; Osterwalder, J.; Greber, T., Surface Trapping of Atoms and Molecules with Dipole Rings. *Science* **319**, 1824-1826 (2008).
30. Joshi, S.; Ecija, D.; Koitz, R.; Iannuzzi, M.; Seitsonen, A. P.; Hutter, J.; Sachdev, H.; Vijayaraghavan, S.; Bischoff, F.; Seufert, K.; Barth, J. V.; Auwärter, W., Boron Nitride on Cu(111): An Electronically Corrugated Monolayer. *Nano Lett.* **12**, 5821-5828 (2012).
31. Schulz, F.; Drost, R.; Hämäläinen, S. K.; Demonchaux, T.; Seitsonen, A. P.; Liljeroth, P., Epitaxial hexagonal boron nitride on Ir(111): A work function template. *Phys. Rev. B* **89**, 235429 (2014).
32. Zhang, C.; Chen, Y.; Johnson, A.; Li, M.-Y.; Li, L.-J.; Mende, P. C.; Feenstra, R. M.; Shih, C.-K., Probing Critical Point Energies of Transition Metal Dichalcogenides: Surprising Indirect Gap of Single Layer WSe2. *Nano Lett.* **15**, 6494-6500 (2015).
33. Sutter, P.; Lahiri, J.; Albrecht, P.; Sutter, E., Chemical Vapor Deposition and Etching of High-Quality Monolayer Hexagonal Boron Nitride Films. *ACS Nano* **5**, 7303-7309 (2011).
34. Laskowski, R.; Blaha, P., Ab initio study of h-BN nanomeshes on Ru(001), Rh(111), and Pt(111). *Phys. Rev. B* **81**, 075418 (2010).
35. Brugger, T.; Günther, S.; Wang, B.; Dil, J. H.; Bocquet, M.-L.; Osterwalder, J.; Wintterlin, J.; Greber, T., Comparison of electronic structure and template function of single-layer graphene and a hexagonal boron nitride nanomesh on Ru(0001). *Phys. Rev. B* **79**, 045407 (2009).

36. Gómez Díaz, J.; Ding, Y.; Koitz, R.; Seitsonen, A. P.; Iannuzzi, M.; Hutter, J., Hexagonal boron nitride on transition metal surfaces. *Theor. Chem. Acc.* **132**, 1-17 (2013).

37. Kresse, G.; Furthmüller, J., Efficiency of ab-initio total energy calculations for metals and semiconductors using a plane-wave basis set. *Comput. Mater. Sci.* **6**, 15–50 (1996)

38. Blöchl, P. E., Projector augmented-wave method. *Phys. Rev. B* **50**, 17953-17979 (1994).

39. Perdew, J. P.; Burke, K.; Ernzerhof, M., Generalized Gradient Approximation Made Simple. *Phys. Rev. Lett.* **77**, 3865-3868 (1996).

**Acknowledgements:**

This research was supported with grants from the Welch Foundation (F- 1672), and the US National Science Foundation (DMR-1306878, EFMA-1542747). C.Z. also acknowledge the National Basic Research Program of China (Grant No. 2014CB921102).

**Author contributions:**

Q.Z. and Y.C. carried out the growth of heterostructures and STM/S measurements. C.Z., Y.C., and C.D.Z. carried out the data analysis. C.R.P and M.Y.C carried out the first-principles calculation. C.K.S. and C.Z. advised on the experiment and provided input on the data analysis. Q.Z., Y.C., and C.K.S. wrote the paper with input from other co-authors.

**Additional information**

**Competing financial interests**

The authors declare no competing financial interests.

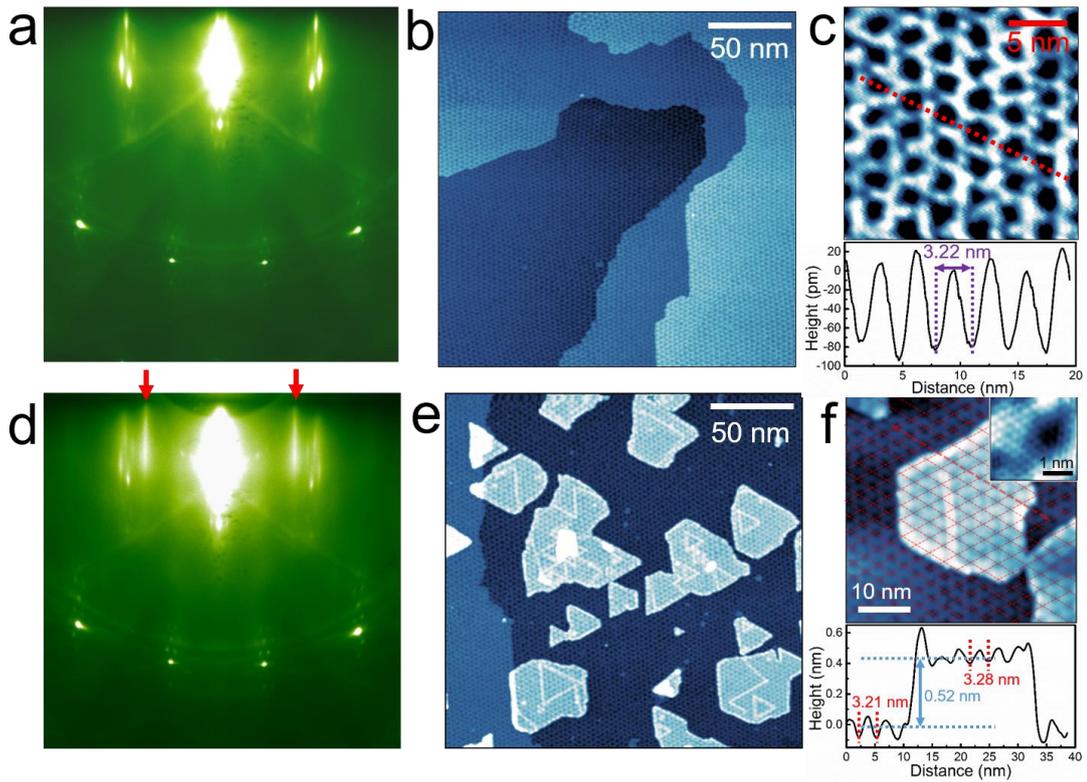

**Figure 1 | RHEED and STM characterizations of hBN/Ru(0001) and MoSe$_2$/hBN/Ru(0001).** (a) RHEED pattern of epitaxial single layer hBN on the Ru(0001) substrate. (b) Large-scale STM image (-2 V, 5 pA; 215 nm × 215 nm) shows the uniform high-quality single-layer hBN on Ru(0001). (C) The upper panel is the high-resolution constant-current STM image (-2 V, 5 pA, 20 nm × 20 nm) of the h-BN "nanomesh" (moiré pattern), and the lower panel is the line profile along the red dashed line. The periodicity and corrugation of the nanomesh are about 3.2 nm and 0.1 nm, respectively. (d) RHEED pattern of the monolayer MoSe$_2$ thin film grown on hBN/Ru(0001). The sharp and uniform streaks (indicated by red arrows) reflect the successful synthesis of the MoSe$_2$ film. (e) Large-scale STM image (-2 V, 5 pA; 215 nm × 215 nm) shows the successful fabrication of nanoscale MoSe$_2$ islands. The island size ranges from a few nanometers to a hundred nanometers. (f) The top one is the STM image of a typical MoSe$_2$ island, and the inset reveals the atomic resolution of the MoSe$_2$ layer. The bottom one is the line profile along the red dashed line. The corrugation of the MoSe$_2$ surface is completely in phase with that of the underlying hBN, indicating that the periodic structure on the island is actually a transparent pattern

from the nanomesh through the single-layer MoSe$_2$, instead of a moiré pattern formed by MoSe$_2$ and the underlying hBN.

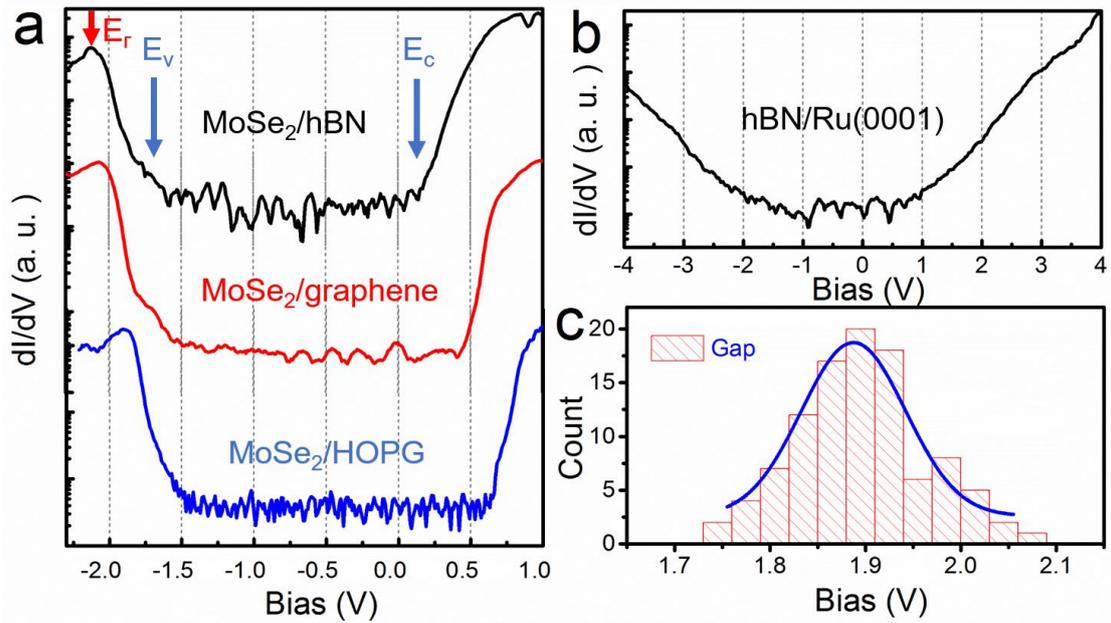

**Figure 2 | Tunneling spectroscopy of MoSe$_2$ and hBN/Ru(0001). (a)** Logarithm of dI/dV spectra for SL-MoSe$_2$ grown on different substrates. MoSe$_2$/hBN/Ru(0001), MoSe$_2$/graphene/SiC and MoSe$_2$/HOPG are shown in black, red and blue, respectively. The MoSe$_2$ layer on hBN has a smaller quasi-particle band gap (by about 0.25 eV) than that of MoSe$_2$ on graphene or graphite substrates. **(b)** Logarithmic dI/dV of hBN/Ru(0001) shows the remanent metallic characteristics, due to the strong interaction between single layer hBN and Ru(0001). **(c)** Statistical distribution (from 102 individual dI/dV spectrum) of the quasi-particle band gap measured for MoSe$_2$/hBN. Mean value: 1.90 eV. Standard deviation: 0.07 eV.

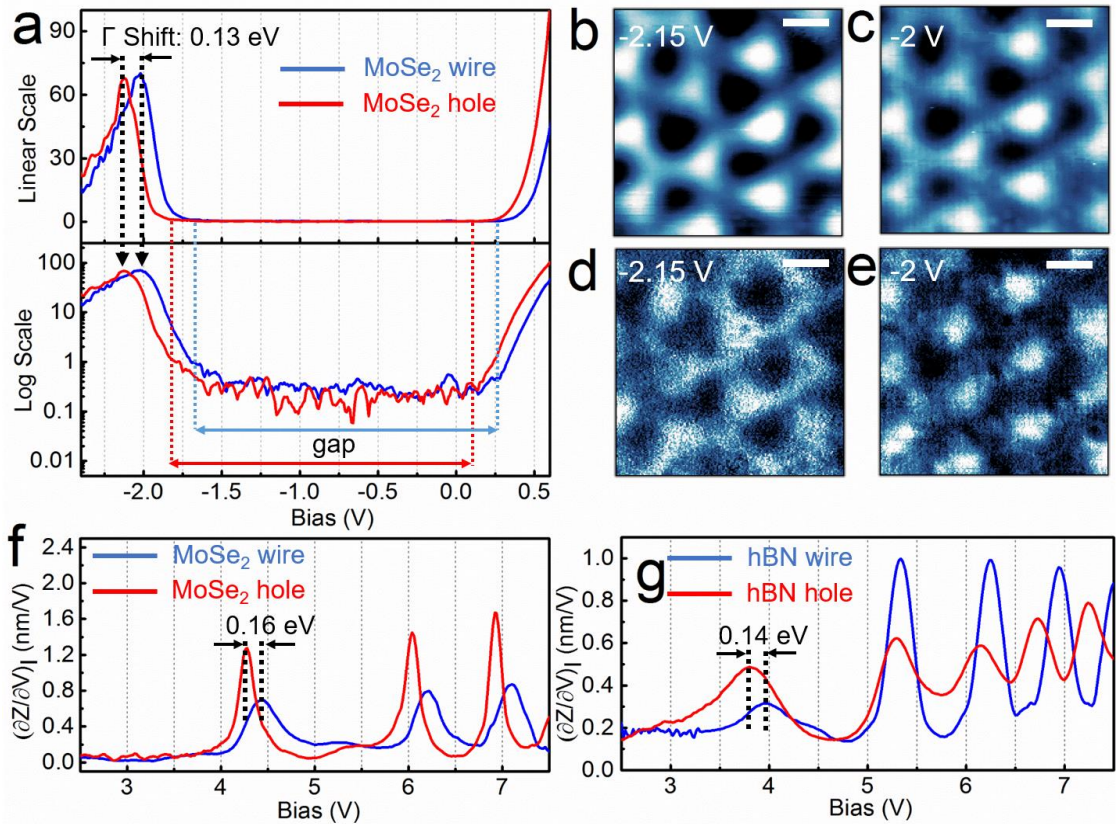

**Figure 3 | STM images and the tunneling spectra of MoSe₂ taken from hole and wire locations. (a)** The dI/dV spectrum taken on the SL-MoSe₂ flake. The tunneling conductance dI/dV (with arbitrary unit) is plotted in both the linear scale (upper panel) and the logarithmic scale (lower panel). The black dashed arrows indicate the Γ points and we observed a rigid shift of the whole band structures (by about 0.13 eV) on hole and wire locations. **(b,c)** Topography for the corrugated SL-MoSe₂. **(d,e)** Corresponding dI/dV images for **(b)** and **(c)**, respectively. Scale bar: 2 nm. **(f)** FER spectroscopy measured on the MoSe₂ wire (blue) and MoSe₂ hole (red). **(g)** FER spectroscopy measured on the hBN wire (blue) and hBN hole (red).

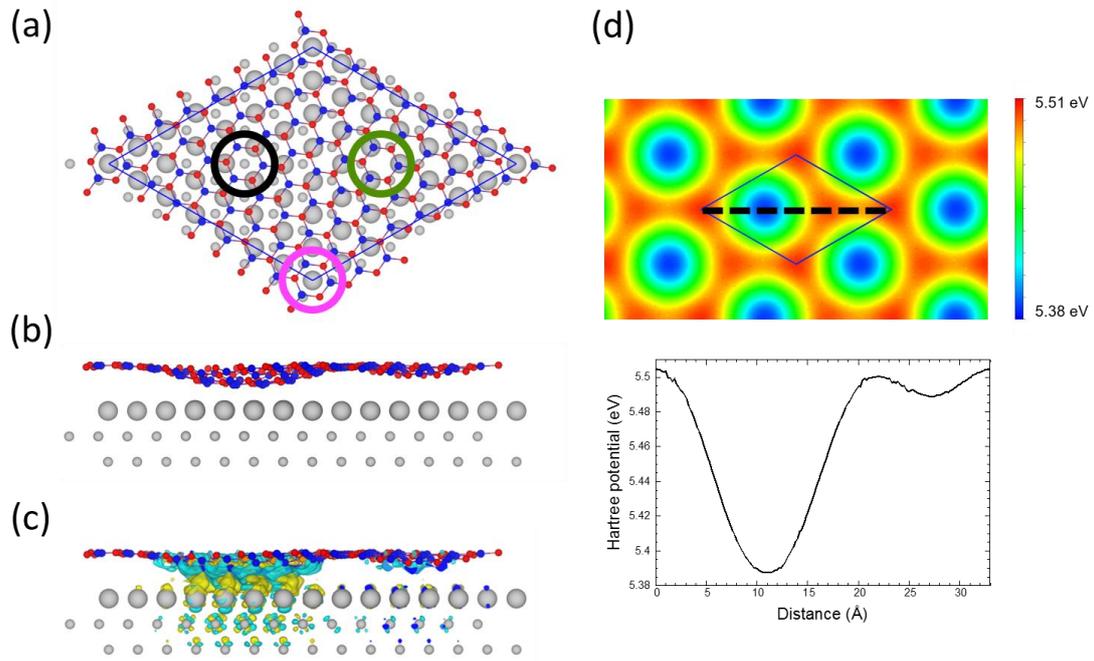

**Figure 4 | First-principles calculations for the electronic structures of hBN/Ru(0001). (a)** Top view and **(b)** side view of the atomic structure of √57 x √57 h-BN on 7 x 7 Ru(0001). The red, blue, and gray spheres are N, B, and Ru atoms, respectively. The black, pink, and green circles indicate the regions with N atoms at the top, fcc, and hcp sites, respectively, with respect to the Ru(0001) substrate. The distance between the h-BN and the surface Ru layers is about 3.7 Å in the regions indicated by both the green and pink circles and 2.15 Å in the region of the black circle with a maximal height difference of about 1.6 Å for the h-BN layer. **(c)** Charge density difference induced by the interaction, with the yellow (blue) isosurfaces indicating an increase (decrease) in the charge density. **(d)** Calculated electrostatic potential variations at the height of 4.9 Å above the h-BN layer, which corresponds to average of the interlayer distances of h-BN and $MoSe_2$. The profile along the dotted black line is also shown.

# Supplementary Figures

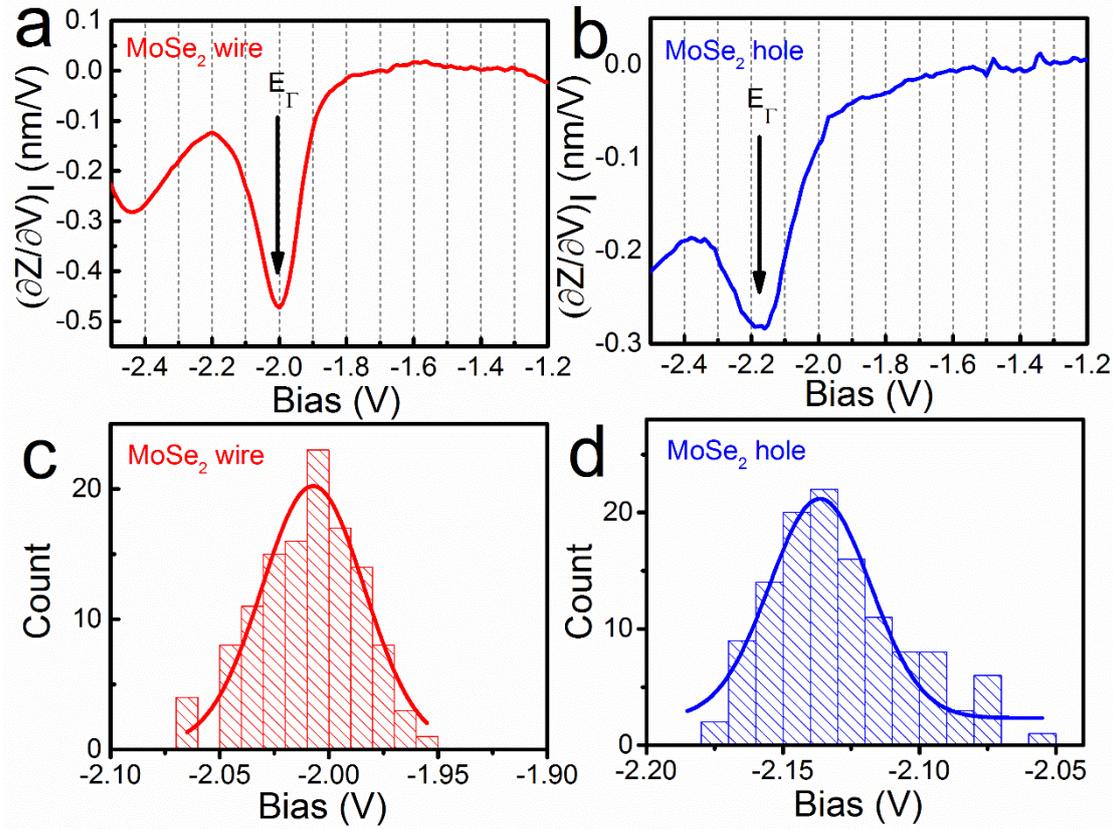

**Supplementary Figure 1 | (∂Z/∂V)_I spectra and statistical distributions of the Γ points. (a,b)** Individual (∂Z/∂V)_I spectra taken from MoSe₂ hole and MoSe₂ wire locations, respectively. The black arrows indicated the energy locations of the Γ points[1,2]. **(c,d)** Statistical distributions for Γ points of wire and hole, respectively (based on 120 individual (∂Z/∂V)_I spectrum). $\Gamma_V^H = -2.14 \pm 0.03$ eV, $\Gamma_V^W = -2.01 \pm 0.02$ eV

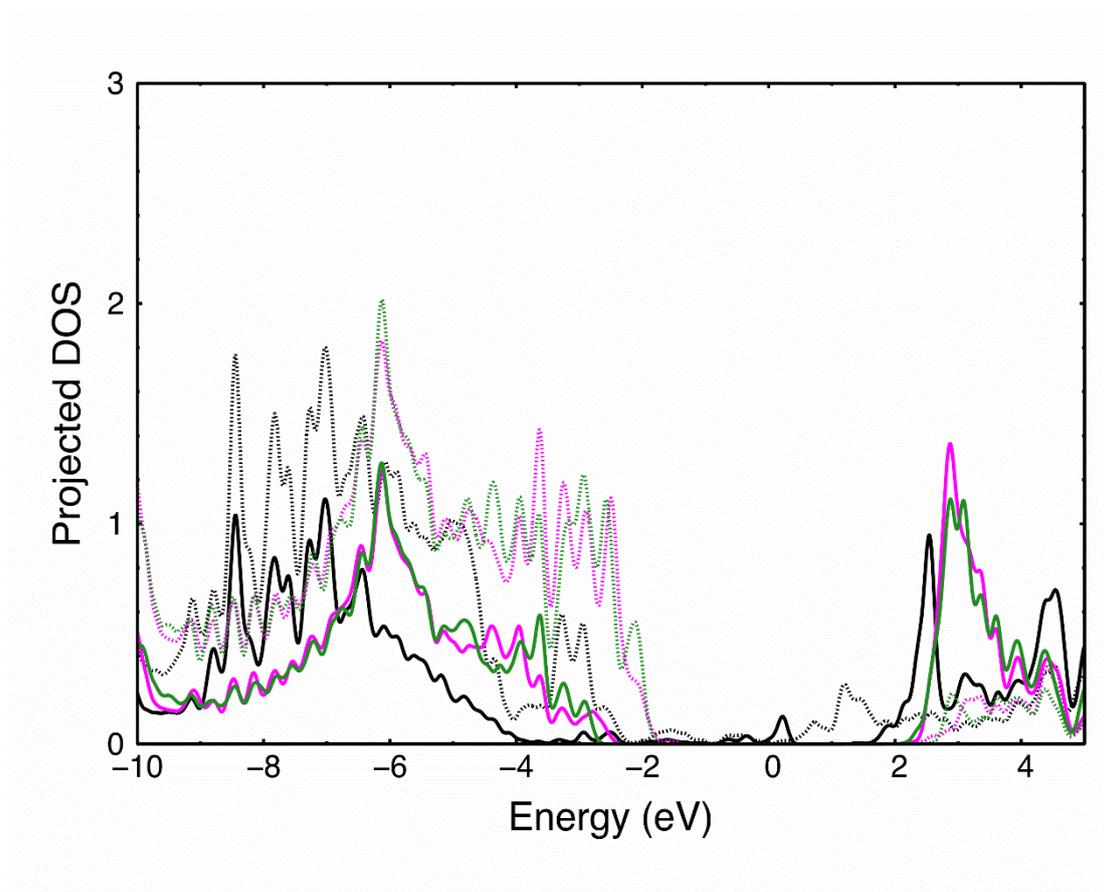

**Supplementary Figure 2** | Projected density of states on the p orbitals of B (solid lines) and N (dotted lines) atoms in the regions indicated by the black, green, and purple circles in Figure 4a. The Fermi level is set as the zero energy.

**References:**
1.  Zhang, C.; Chen, Y.; Johnson, A.; Li, M.-Y.; Li, L.-J.; Mende, P. C.; Feenstra, R. M.; Shih, C.-K., Probing Critical Point Energies of Transition Metal Dichalcogenides: Surprising Indirect Gap of Single Layer WSe2. *Nano Lett.* **15**, 6494-6500 (2015).
2.  Zhang, C.; Chen, Y.; Huang, J.-K.; Wu, X.; Li, L.-J.; Yao, W.; Tersoff, J.; Shih, C.-K., Visualizing band offsets and edge states in bilayer-monolayer transition metal dichalcogenides lateral heterojunction. *Nat. Commun.* **7**, 10349 (2016).